\documentclass[aps,pra,amssymb,amsmath,eqsecnum,nofootinbib]{revtex4}
\usepackage{graphicx}

\newtheorem{definition}{Definition}[section]
\newtheorem{theorem}{Theorem}[section]

\newtheorem{remark}{Remark}[section]

\newenvironment{hypothesis}{HP: \begin{center}} {\end{center}}
\newenvironment{thesis}{TH: \begin{center}} {\end{center}}
\newenvironment{proof}{\begin{center}PROOF: \end{center}} {$ \blacksquare $}
\newtheorem{example}{Example}[section]
\begin{document}
\title{Magnetic Bottles on Riemann Surfaces}
\author{Gavriel Segre}
\begin{abstract}
Yves Colin de Verdiere's quantization formalism of magnetic bottles on Riemann surfaces of non null genus is shown to be affected, owing to the Homotopy Superselection Rule, by the phenomenon of the existence of multiple inequivalent quantizations mathematically analogous to the phenomenon of the existence of multiple inequivalent prequantizations of a multiply-connected symplectic manifold in the framework of Souriau-Kostant's Geometric Quantization.
\end{abstract}
\maketitle
\newpage
\tableofcontents
\newpage
\section{Acknowledgements}
I would like to thank Giovanni Jona-Lasinio and Vittorio de Alfaro for many useful suggestions.

Of course nobody among them has any responsibility a to any (eventual) mistake contained in these pages.
\newpage
\section{Introduction}
This paper contains the main result of my  first degree thesis that I defended in July 1997 at the Physics Department of the Universit\'{a} La Sapienza
of Rome having as tutor my teacher and friend Giovanni Jona-Lasinio.

In 1986 the french mathematical physicist Yves Colin de Verdiere \cite{de-Verdiere-86}, following previous analyses about magnetic Schr\"{o}dinger  operators performed by Barry Simon and coworkers \cite{Cycon-Froese-Kirsch-Simon-87}, \cite{Erdos-07}, introduced a mathematical formalism devoted to the quantization of particles having as configuration space a riemannian manifold \cite{Nakahara-03} and subjected to a magnetic field orthogonal to such a manifold.

Such a formalism is very similar to the procedure of prequantization of Souriau-Kostant's Geometric Quantization \cite{Woodhouse-91}, \cite{Souriau-97}.

In the case in which the underlying simplectic manifold is multiply-connected the phenomenon of \emph{Homotopy Superselection Rule} (i.e. the phenomenon discovered at the end of the sixthes and the beginning of the seventhes by Larry Schulman and  Cecile Morette De Witt according to which there exist inequivalent quantizations of dynamical systems having a multiply-connected configuration space; see for instance the $ 23^{th} $ chapter of \cite{Schulman-81}, the $ 7^{th}$ chapter of \cite{Rivers-87} and the $ 8^{th}$ chapter of \cite{Cartier-De-Witt-Morette-06} as to its implementation, at different levels of mathematical rigor, in the path-integration's formulation, as well as the $ 8^{th} $ chapter of \cite{Balachandran-Marmo-Skagerstam-Stern-91}, the $3^{th}$ chapter of \cite{Morandi-92} and the section $ 6.8 $ of \cite{Strocchi-05b} for its formulation in the operatorial formalism) appears in such a formalism as the existence of inequivalent prequantizations.

This fact has led us to infer that the same phenomenon occurs in de Verdiere's quantization of multiply-connected magnetic bottles, a fact that is here deeply analyzed.

\newpage
\section{Homotopy Superselection Rule} \label{sec:Homotopy Superselection Rule}
The phenomenon of \emph{Homotopy Superselection Rule} (cfr. the previously mentioned literature) consists in the fact that in presence of a multiply connected configuration space, let's call it M, a new topological superselection rule arises, the involved superselection charge taking values in $ Hom ( H_{1}(M , \mathbb{Z}) , U(1)) $ where we remind that according to the Hurewicz Isomorphism:
\begin{equation}
    H_{1}(M , \mathbb{Z} ) \; = \; \frac{\pi_{1} (M)}{[ \pi_{1} (M),\pi_{1} (M)]}
\end{equation}
where:
\begin{equation}
    [ G , G ] \; := \; \{ x \cdot y \cdot x^{-1} \cdot y^{- 1} \; \; x,y \in G \}
\end{equation}
is the commutator subgroup of the group G.

Such a phenomenon appears in physically very different contexts going from the $ \theta$-angle of Yang-Mills Quantum Field Theories (see for instance the $ 10^{th}$ chapter of \cite{Ryder-96} or the section $23.6$ of \cite{Weinberg-96}) to the magnetic flux of the solenoid involved in the Aharonov-Bohm effect and to the fractional statistic of a quantum system of identical particles living on the plane (see for instance \cite{Wilczek-90} and \cite{Lerda-92}).

Hence, in such a situation, the Hilbert space decomposes in the direct sum of superselection sectors:
\begin{equation}
  \mathcal{H} \; = \; \, \oplus_{\theta \in Hom ( H_{1}(M, \mathbb{Z}) , U(1))} \, \mathcal{H}_{\theta}
\end{equation}
where of course the family of superselection operators is made of the operators of orthogonal projections over the superselections sectors.

\newpage
\section{The Kato operator of a closed symmetric form}
Given an Hilbert space $ \mathcal{H}$:
\begin{definition}
\end{definition}
\emph{symmetric form on $ \mathcal{H}$:}

$ (E , D(E) ) \; := E : D(E) \times   D(E) \mapsto \mathbb{R} $ such that:
\begin{enumerate}
  \item D(E) is a linear subspace dense in  $ \mathcal{H}$:
  \item E is linear
  \item
  \begin{equation}
    E( \psi , \psi ) \; \geq \; 0 \; \; \forall \psi \in D(E)
  \end{equation}
\end{enumerate}

Given a symmetric form $ (E , D(E))$ one can derive from it the new symmetric form $ (E_{1} , D(E_{1})) $ where:
\begin{equation}
    D(E_{1}) \; := \; D(E)
\end{equation}
\begin{equation}
    E_{1}(f,g) \; := \; E( f, g) \, + \, ( f , g)
\end{equation}

We will say that:
\begin{definition}
\end{definition}
\emph{$ (E , D(E) )$ is closed:}
\begin{center}
    $ D(E) $ is complete with respect to the norm $ \| \cdot \|_{1} := \sqrt{E_{1}( \cdot , \cdot)}$
\end{center}
\begin{definition}
\end{definition}
\emph{$ (E , D(E) )$ is closable:}
\begin{center}
    it has closed extension
\end{center}

Let us assume that  $ (E , D(E))$ is closable:
\begin{definition}
\end{definition}
\emph{closure of $ (E , D(E) )$:}
\begin{center}
    $ \overline{(E , D(E) )} \; := \; $ the smallest closed extension of $ (E , D(E) )$
\end{center}

Then \cite{Reed-Simon-80}, \cite{Reed-Simon-75}:
\begin{theorem} \label{th:Kato operator}
\end{theorem}
\emph{Theorem about Kato operator:}

\begin{hypothesis}
\end{hypothesis}
\begin{center}
    $ ( E, D(E) ) $ closed symmetric form
\end{center}
\begin{thesis}
\end{thesis}
\begin{center}
    $  \exists  \,! \,  K[ ( E, D(E) ) ] \, \text{positive, self-adjoint operator} \; : \; $
\end{center}
\begin{equation}
    D(E) = D( \sqrt{ K[ ( E, D(E) ) ]})
\end{equation}
\begin{equation}
    E( f,g) \, = \; ( \sqrt{ K[ ( E, D(E) ) ]} f , \sqrt{ K[ ( E, D(E) ) ]} g )
\end{equation}

The positive self-adjoint operator $ K[ ( E, D(E) ) ] $ associated by the theorem \ref{th:Kato operator} to a closed symmetric form $ ( E, D(E) ) $ is called the \emph{Kato operator} of such a form.
\newpage
\section{Quantization of a free particle on a simply-connected Riemannian manifold}
Let us consider the classical dynamical system consisting in a free particle of unary mass, i.e. having action $ S[q(t)] := \int dt \frac{| \dot{q}|_{g}^{2}}{2} $,  living on a simply connected riemannian manifold $ ( M , g = g_{\mu \nu } d x^{\mu} \otimes d x^{\nu} ) $.

There exists a long debate in the literature about which is the quantum dynamical system obtained quantizing such a classical dynamical system (see for instance the $ 24^{th}$ chapter of \cite{Schulman-81}, the $ 9^{th} $ chapter of \cite{Woodhouse-91}, the $ 3^{th} $ chapter of \cite{Zinn-Justin-93}, the $10^{th}$ and $ 11^{th}$ chapter of \cite{Kleinert-95} and the $ 15^{th} $ chapter of \cite{De-Witt-03}).

Most of the authors concord that the quantum hamiltonian of such a system is of the form:
\begin{equation}
    D ( \hat{H} ) \; = \; S^{2}(M)
\end{equation}
\begin{equation}
    \hat{H} \; = \; \frac{1}{2} \Delta_{g} \, + \, c R_{g}
\end{equation}
where $ S^{k}(M) $ is the $k^{th}$ Sobolev space over M, $\Delta_{g} = - \frac{1}{\sqrt{| g |}} \partial_{\mu} ( \sqrt{| g |} g^{\mu \nu} \partial_{\nu} f)$ is the Laplace-Beltrami operator over $ ( M ,g)$ (where of course $ \partial_{\mu} := \frac{\partial}{ \partial x^{\mu} } $ while $ g = det(g_{\mu \nu} ) $) and $ R_{g}$ is the scalar curvature of the metric g \cite{Nakahara-03}.

As to the value of the constant c the more palatable proposals are $ c=0 $ (according to Cecile Morette De-Witt), $ c = \frac{1}{6}$ (in conformity with the $1^{th}$ order term in the asymptotic expansion of the heat-kernel \cite{Gilkey-95} over $ ( M , g) $, $ c = \frac{1}{12}$ (as according to the first Bryce De-Witt), $ c= \frac{1}{8}$ (as according to the last  Bryce De-Witt) and various other alternatives (someway related to the fact that, for $ D \geq 2$, $ \hat{H}$ is  invariant under conformal transformation if and only if $ c = \frac{D-2}{4(D-1)} $).

We will assume that the correct quantum Hamiltonian is:
\begin{equation}
    ( D( \hat{H}) , \hat{H} ) \; = \; K [  \overline{( D(E) := C_{0}^{\infty}(M) , E ( \psi_{1} ,\psi_{2}) := \int_{M} d \mu_{g} < \nabla_{g} \psi_{1}  , \nabla_{g} \psi_{2} >_{g}  )} ]
\end{equation}
(where $ d \mu_{g} := \sqrt{|g|} d x^{1} \cdots dx^{dim(M)}$ is the invariant measure over $ ( M , g) $ and $ \nabla_{g}$ is the Levi-Civita connection of $ ( M , g) $) and hence that Cecile Morette De-Witt is right:
\begin{equation}
    c \; = \; 0
\end{equation}
\newpage
\section{Quantization of a free particle on a Riemann  surface of genus $ g \neq 0$}
Let us now  consider the classical dynamical system consisting in a free particle of unary mass, i.e. having action $ S[q(t)] := \int dt \frac{| \dot{q}|_{g}^{2}}{2} $,  living on a Riemann surface  $ ( \Sigma_{g}, g = g_{\mu \nu } d x^{\mu} \otimes d x^{\nu} ) $ of genus $ g \neq 0$.

\begin{theorem} \label{homotopic lattice}
\end{theorem}
\begin{equation}
    H_{1} ( \Sigma_{g} , \mathbb{Z} ) \; = \; \mathbb{Z}^{2 g}
\end{equation}
\begin{proof}
The Theorem of Classification of Bidimensional Manifolds \cite{Dubrovin-Novikov-Fomenko-92}, \cite{Schlichenmaier-07} guarantees that there exists a triangulation of $ \Sigma_{g} $ to a simplicial complex, called the normal topological form of $ \Sigma_{g}$, that may be codified through the word in the 1-simplexes:
\begin{equation}
    Ntf( \Sigma_{g}) \; = \; \prod_{k=1}^{g} a_{k} b_{k} a_{k}^{-1} b_{k}^{-1}
\end{equation}

To it there corresponds the system of generators of the $1 ^{th}$ homotopy group of $ \Sigma_{g} $ $ \{ a_{1}, b_{1} , \cdots, a_{g}, b_{g} \}$.

Clearly they have to satisfy the relation:
\begin{equation} \label{eq:relation linking the generators of the fundamental group}
     \prod_{k=1}^{g} a_{k} b_{k} a_{k}^{-1} b_{k}^{-1} \; = \; 1
\end{equation}
since the loop made of the boundary of the normal topological form is contractible over $ \Sigma_{g}$.

The equation \ref{eq:relation linking the generators of the fundamental group} implies, in particular, that for $ g > 1 $ $ \pi_{1} ( \Sigma_{g}) $ is not abelian since loops winding different handles don't commute.

The thesis immediately follows.
\end{proof}

\bigskip

The theorem \ref{homotopic lattice} implies, in particular, that the torsion subgroup of the first simplicial homology of $ \Sigma_{g} $ is null.

We will call $ \Gamma := H_{1} ( \Sigma_{g} , \mathbb{Z} ) \; = \; \mathbb{Z}^{2 g} $ the \emph{homotopic lattice}.

\begin{theorem} \label{th:flux torus}
\end{theorem}
\begin{equation}
    Hom ( H_{1} ( \Sigma_{g} , \mathbb{Z} ) , U(1) ) \; = \; T^{2 g } \; = \frac{\mathbb{R}^{2g}}{2 \pi \mathbb{Z}^{2 g}} \; = \; \times_{k=1}^{2g} S^{1}
\end{equation}
\begin{proof}
Let us introduce an intersection product $ \cdot :  H_{1} ( \Sigma_{g} , \mathbb{Z} ) \times  H_{1} ( \Sigma_{g} , \mathbb{Z} ) \mapsto \mathbb{Z} $ by defining it on the homology 1-cycles of its normal topological form by:
\begin{equation}
    a_{i} \cdot a_{j} \; := \; b_{i} \cdot b_{j} \; := \; 0 \; \; i,j=1,\cdots, g
\end{equation}
\begin{equation}
    a_{i} \cdot b_{j} \; := \; - a_{j} \cdot b_{i} \; := \; \delta_{i j} \; \; i,j=1,\cdots, g
\end{equation}
and imposing that on the generic couple of homology 1-cycles:
\begin{equation}
    c^{A} \; = \; \sum_{k=1}^{g} n_{k}^{A} a_{k} + m_{k}^{A} b_{k} \; \; n_{k}^{A} , m_{k}^{A} \in \mathbb{Z} \, , \, k = 1, \cdots , g
\end{equation}
\begin{equation}
    c^{B} \; = \; \sum_{k=1}^{g} n_{k}^{B} a_{k} + m_{k}^{B} b_{k} \; \; n_{k}^{B} , m_{k}^{B} \in \mathbb{Z} \, , \, k = 1, \cdots , g
\end{equation}
it takes the value:
\begin{equation} \label{eq:intersection  matrix}
    c^{A} \cdot  c^{B} \; = \;
    \left(
    \begin{array}{cc}
     n^{A} & m^{A} \\
     \end{array}
     \right)
     \left(
       \begin{array}{cc}
         0 & I \\
         - I & 0 \\
       \end{array}
     \right)
     \left(
       \begin{array}{c}
         n^{B} \\
         m^{B} \\
       \end{array}
     \right)
\end{equation}
The matrix  $ 2g \times 2g $ at the right hand side of the equation \ref{eq:intersection matrix} is called  intersection matrix of the intersection product
in the basis of homology 1-cycles of its  normal topological form.

Obviously one can choose a different basis.

A basis in which the intersection matrix assumes the form $ \left(
                                                              \begin{array}{cc}
                                                                0 & I \\
                                                                -I  & 0 \\
                                                              \end{array}
                                                            \right) $ is called a canonical basis of homology.

By choosing a canonical basis of homology the thesis easily follows.
\end{proof}

\bigskip

We will call $ \Phi \; = \; Hom ( H_{1} ( \Sigma_{g} , \mathbb{Z} ) , U(1) ) \; = \; T^{2 g } $ the \emph{flux torus}.

The existence of the \emph{Homotopy Superselection Rule} discussed in the section \ref{sec:Homotopy Superselection Rule} implies that the Hilbert space of quantum states
decomposes in the direct integral:
\begin{equation} \label{eq:direct integral decomposition in hopotopic superselection sectors}
    \mathcal{H} \; = \; \int_{\Phi}^{\oplus} \frac{ d \vec{\theta}}{( 2 \pi )^{2 g }} \mathcal{H}_{\vec{\theta}}
\end{equation}
performed with respect to the measure $ d \vec{\theta} $ induced on the flux torus by the immersion in it of the first Brillouin zone of the homotopic lattice endowed with the Lebesgue measure.

\bigskip

\begin{remark}
\end{remark}

It is useful to appreciate how the direct integral \ref{eq:direct integral decomposition in hopotopic superselection sectors} is similar to the one occurring in the Bloch-Floquet theory of Schrodinger operators invariant under the group of translations of vectors belonging to a Bravais lattice \cite{Reed-Simon-78}, \cite{Berezin-Shubin-91}.

This is not a coincidence:

a periodic Schrodinger operator on the real axis may be thought as the lifted to the universal covering space of a Schrodinger operator on the circle.

In an analogous way a Schrodinger operator on $ \mathbb{R}^{D} $ invariant under the group of translations of vectors belonging to a Bravais lattice may be thought as the lifted to the universal covering space of a Schrodinger operator on the D-torus.

This consideration allows to follow a whole quantization strategy, that we won't follow in this paper,  consisting in adopting the Generalized Bloch Theorem \cite{Morandi-92} to determine the spectral properties of the homotopic lifted to the universal covering space projecting only at the end to the original multiply-connected configuration space.

\bigskip

On each homotopy superselection sector the quantum hamiltonian is then given by:
\begin{equation}
    ( D( \hat{H}_{\vec{\theta}}) , \hat{H}_{\vec{\theta}} ) \; = \; K [  \overline{( D(E_{\vec{\theta}}) := C_{0}^{\infty}(\Sigma_{g}) , E_{\vec{\theta}} ( \psi_{1} ,\psi_{2}) := \int_{\Sigma} d \mu_{g} < \nabla_{g}^{\vec{\theta}} \psi_{1}  , \nabla_{g}^{\vec{\theta}} \psi_{2} >_{g}  )} ]
\end{equation}
\newpage
\section{Quantization of a trivial magnetic bottle over a simply-connected riemannian manifold}

Let us start from the following:
\begin{definition}
\end{definition}
\emph{magnetic bottle:}

a couple $ ( ( M , g) , B) $ where:
\begin{itemize}
  \item $ ( M , g) $ is a riemannian manifold
  \item $ B \in Z^{2}(M) $
\end{itemize}

Given a magnetic bottle $ (( M , g) , B)$:
\begin{definition}
\end{definition}
\emph{ $ (( M , g) , B)$ is trivial:}
\begin{center}
    $ B \in B^{2}(M)$
\end{center}

\smallskip

\begin{remark} \label{rem:a trivial magnetic bottles satisfied Weyl integrality condition}
\end{remark}

Given a trivial magnetic bottle  $ (( M , g) , B)$  one has that:
\begin{equation}
   [  \frac{B}{2 \pi} ] \; = \; [ \mathbb{I} ] \; \in \; H^{2} ( M , \mathbb{Z} )
\end{equation}

\smallskip

Let us suppose that M is simply-connected and that the magnetic bottle $ (( M , g) , B)$ is trivial.

Since the 2-form B is exact there exists a 1-form $ A  \in \Omega^{1}(M) $ such that $ B \; = d A $

Introduced the trivial hermitian linear bundle $ L := M \times \mathbb{C} $ (whose hermitian structure is naturally induced by the riemannian structure of $ (M , g)$)
  and the hermitian connection $ D  \; := \; \nabla_{g} - i A $ the quantum hamiltonian obtained quantizing $ (( M , g) , B)$ is:
  \begin{equation}
    ( D( \hat{H}) , \hat{H}) \; := \; K[ \overline{(  D(E) = \Gamma_{0}^{\infty} (L) \, , \,  E ( \psi_{1}, \psi_{2} ) = \int_{M} d \mu_{g} < D \psi_{1} , D \psi_{2}> )  }]
  \end{equation}

\newpage
\section{Quantization of a non-trivial symply-connected magnetic bottle}

Let us now consider the quantization of a non-trivial magnetic bottle $ ((M  , g ) , B ) $ on a symply-connected manifold M.

Since the magnetic field's closed 2-form is not exact $ B \notin B^{2} ( M )$ it cannot be integrated globally.

Anyway it can be integrated locally, i.e. there exists a contractible open covering $ \{ U_{i} \}_{i \in I} $ (where I is a suitable index set) and a collection of 1-forms $  \{ A_{i} \in \Omega^{1} ( U_{i} ) \}_{i \in I} $ such that:
\begin{equation}
    B |_{U_{i}} \; = \; d A_{i} \; \; \forall i \in I
\end{equation}

One can the consider the family of local quantum hamiltonians:
\begin{equation}
    ( D( \hat{H}_{i}) , \hat{H}_{i} ) \; := \; K [ \overline{( D(E) := C_{0}^{\infty} ( U_{i}) \, , \, E ( \psi_{1} , \psi_{2} ) := \int_{M} d \mu_{g} < (\nabla_{g} -i A_{i}) \psi_{1},(\nabla_{g} -i A_{i}) \psi_{2} >_{g}  ) }
\end{equation}

The problem then naturally arises to determine the conditions under which this collection of local quantum hamiltonians may be patched together to define a global quantum hamiltonian
acting on the global sections of a suitable fibre bundle.

Let us then introduce, with this regard, the following:
\begin{definition}
\end{definition}
\emph{$1^{th}$ Weyl integrality condition}:
\begin{equation}
    \int_{S} B \; \in \; 2 \pi \mathbb{Z} \; \; \forall \text{ closed oriented 2-surface S in M}
\end{equation}

\begin{definition}
\end{definition}
\emph{$ ((M , g), B )$ is quantizable:}
\begin{center}
    $ \exists \; L \stackrel{\pi}{\rightarrow} M $ hermitian linear bundle endowed with an hermitian connection $ \nabla $ on it such that $ F_{\nabla} \; := \; \nabla \circ \nabla \; = \; B $
\end{center}

Then:
\begin{theorem} \label{th:theorem on the first Weyl integrality condition}
\end{theorem}
\emph{Theorem on the $ 1^{th} $  Weyl integrality condition}
\begin{center}
    $ ((M , g), B )$ is quantizable $ \Leftrightarrow $ the $ 1^{th} $  Weyl integrality condition holds
\end{center}
\begin{proof}
\begin{enumerate}
  \item Let us proof that if $ ((M , g), B )$ is quantizable then  the $ 1^{th} $ Weyl integrality condition holds.

  By hypothesis  $  ((M , g), B )$ admits an hermitian linear bundle $ L \stackrel{\pi}{\rightarrow}  M $ endowed with an hermitian connection $ \nabla $ such that $ F_{\nabla} := \nabla \circ \nabla = B$.

  Given a loop $ \gamma \in LOOPS(M) $ let us consider a surface $ \Sigma_{1}$ such that $ \gamma = \partial \Sigma_{1} $.

  The holonomy of $  \nabla $ along $ \gamma $ is an automorphism of the fibre $ L_{q} , q \in Im( \gamma )$, independent from the choice of the point q along the trajectory of $ \gamma $, such that:
  \begin{eqnarray}
    Hol_{\gamma}(\nabla ) \; &=&  \; L_{q} \mapsto L_{q} \\
    Hol_{\gamma}( \nabla ) \; &=& \; \text{multiplication by} \; \exp ( i \int_{\Sigma_{1}} B )
  \end{eqnarray}

  Let us now consider a different surface $ \Sigma_{2} \neq \Sigma_{1} $ such that  $ \gamma = \partial \Sigma_{2} $

  Since:
\begin{equation}
    Hol_{\gamma}( \nabla ) \; = \; \exp ( i \int_{\Sigma_{1}} B ) \; = \; \exp ( i \int_{\Sigma_{2}} B )
\end{equation}
it follows that:
\begin{equation}
  \int_{\Sigma_{1} \cup \Sigma_{2} } B \; \in \; 2 \pi \mathbb{Z}
\end{equation}
Since every closed surface may be constructed in this way the quantization of the flux along it follows.

    \item Let us now proof that if the $ 1^{th} $  Weyl integrality condition holds then  $ ((M , g), B )$ is quantizable.

    Let us, first of all, fix a base point $q_{0}$ on M.

    Let us then consider triples of the form $ (q, z , \gamma ) $ such that $ q \in M $, $ z \in \mathbb{C} $ and $ \gamma $ is an oriented path joining $ q_{0} $ and q.

    Let us then define on such triples the following equivalence relation:
\begin{equation}
        ( q_{1} , z_{1} , \gamma_{1} ) \, \sim \,  ( q_{2} , z_{2} , \gamma_{2} ) \; := \; q_{1} = q_{2} \, \wedge \,   z_{2} = z_{1} \exp ( i \int_{\Sigma} B )
\end{equation}
    where $ \Sigma $ is a surface having as boundary $ \gamma := \gamma_{1} + ( - \gamma_{2} ) $ that does exists owing to the fact that $ \pi_{1} ( M ) = 0 $ and hence every 1-cycle on M is also a 1-boundary.

    Let us then  consider the quotient set of the triples of the described form with respect to the introduced equivalence relation:
    \begin{equation}
        L \; := \; \{ [ q , z , \gamma ]  \}
    \end{equation}

    L is the total space of a linear bundle $ L \stackrel{\pi}{\rightarrow} M $ defined in the following way:
    \begin{itemize}
      \item addition on the fibres:
      \begin{equation}
        [ ( q, z_{1} , \gamma ) ] \, + \,  [ ( q, z_{2} , \gamma ) ] \; = \; [ ( q, z_{1} + z_{2} , \gamma ) ]
      \end{equation}
      \item scalar multiplication:
      \begin{equation}
        c [ ( q , z , \gamma ) ] \; = \; [ q , c \, z , \gamma ]
      \end{equation}
      \item local trivializations:

      let us consider a a simply connected open U over M, a local potential over U of the magnetic field:
      \begin{equation}
        A \in \Omega^{1} (U) \; = \; B|_{U} = d A
      \end{equation}
      a point $ q_{1} \in U $ and a path $ \gamma_{0} $ joining $ q_{0} $ and $ q_{1} $; let us then define the local section  $ \psi \in \Gamma ( U , L ) $ of $ L \stackrel{\pi}{\rightarrow} M $ defined as:
      \begin{equation}
        \psi (q) \; := \; [ ( q , \exp ( - i \int_{\gamma_{1}} A , \gamma ) ) ]
      \end{equation}
      where $ \gamma_{1} $ is a path joining $ q_{1} $ and q while $ \gamma := \gamma_{0} + \gamma_{1} $
    \end{itemize}
    By construction $ \psi $ is independent from $ \gamma_{1} $. A different choice of $ \gamma_{0} $ or of $ q_{1} $, furthermore, alterates $ \psi $ by a global phase $ \psi \mapsto \exp ( i \theta ) \psi \, , \, \theta \in [ 0 , 2 \pi ) $ while the choice of a different local potential over U of the magnetic field, corresponding  to the ansatz:
    \begin{equation}
        A \; \mapsto \; A + d u
    \end{equation}
    where $ u \in C^{\infty}(U) $ is such that: $ u( q_{1} ) = 0 $, alterates $ \psi $ by the local phase:
    \begin{equation}
        \psi \; \mapsto \; \exp ( - i u ) \psi
    \end{equation}

    Repeating such a construction for every chart $ U_{i} $ of an open covering of M it results defined an atlas of local trivializations.
\end{enumerate}

The $ \{  A_{i} \} $, furthermore, are the local potentials of a connection $ \nabla $ whose curvature is equal to the magnetic field B.

The condition:
\begin{equation}
    < \psi , \psi > \; = \; 1
\end{equation}
determinates, finally, an hermitian structure over $ L \stackrel{\pi}{\rightarrow}  M $ with respect to which $ \nabla $ is an hermitian connection
\end{proof}

\bigskip

Let us then suppose that the $ 1^{th} $ Weyl integrality condition holds.

Given the hermitian linear bundle $  \; L \stackrel{\pi}{\rightarrow} M $  endowed with the hermitian connection $ \nabla $ on it such that $ F_{\nabla} \; := \; \nabla \circ \nabla \; = \; B $ whose existence is assured by the theorem \ref{th:theorem on the first Weyl integrality condition} the quantum hamiltonian is simply:
\begin{equation}
    ( D ( \hat{H} ) , \hat{H} ) \; := \; K [  \overline{ ( D(E) := \Gamma_{0}^{\infty} (L) \, , \, E ( \psi_{1} , \psi_{2} ) := \int_{M} d \mu_{g} < \nabla \psi_{1} , \nabla \psi_{2} >) }
\end{equation}

\bigskip

\begin{example}
\end{example}
Let us consider the important particular case in which $ ( M = S^{2} , g = d \theta \otimes d \theta + \sin^{2} \theta d \phi \otimes d \phi $)  is the 2-sphere endowed by the induced metric owed to its embedding in the three-dimensional euclidean space $ ( \mathbb{R}^{3} , \delta = \delta_{\mu \nu} d x^{\mu} \otimes d x^{\nu}) $.

Since the only closed 2-surface in the 2-sphere is the 2-sphere itself,  the $ 1^{th} $  Weyl integrality condition reduces to the condition of quantization of the total flux of the magnetic field along $ S^{2}$, i.e.:
\begin{equation}
    \int_{S^{2}} B \; \in \; 2 \pi \mathbb{Z}
\end{equation}

\newpage
\section{Quantization of a non-trivial multiply-connected magnetic bottle: quantizability}

We have seen in the proof of the theorem \ref{th:theorem on the first Weyl integrality condition} that the hypothesis $ \pi_{1} (M) \, = \, 0 $ played a key role by determining that the 1-cycle determined by two different paths joining two fixed points is also a 1-boundary.

In the case in which $ \pi_{1} (M ) \; \neq \; 0 $, contrary, the $ 1^{th} $ Weyl integrality condition is no more sufficient to guarantee the quantizability of a magnetic bottle $ ( ( M ,g ), B ) $ and it is necessary to introduce the following:
\begin{definition}
\end{definition}
\emph{$ 2^{th}$ Weyl integrality condition}:
\begin{equation}
    [ \frac{B}{2 \pi} ] \; \in \; H^{2} ( M , \mathbb{Z} )
\end{equation}

as it is stated by the following:
\begin{theorem} \label{th:theorem on the second Weyl integrality condition}
\end{theorem}
\emph{Theorem on the $ 2^{th} $  Weyl integrality condition}
\begin{center}
    $ ((M , g), B )$ is quantizable $ \Leftrightarrow $ the $ 2^{th} $  Weyl integrality condition holds
\end{center}
\begin{proof}
\begin{enumerate}
  \item Let us prove that if the $ 2^{th} $  Weyl integrality condition holds then $ ( ( M , g) , B ) $ is quantizable.

  The $ 2^{th} $  Weyl integrality condition implies that there exist a triple $ ( \{ U _{i} \} , \{ A_{i} \} , \{ u_{j k} \})  $ such that:
  \begin{itemize}
    \item $ \{ U_{i} \} $ is a contractible open cover of M
    \item $ \{ A_{i} \} $ is a collection of local potentials of the magnetic field:
    \begin{equation}
        A_{i} \in \Omega^{1}( U_{i} ) \; : \; B|_{U_{i}} \, =  \, d A_{i}
    \end{equation}

  \item $ \{ u_{j k} \} $ is a collection of maps $ u_{j k} \in C^{\infty} ( U_{j} \cap U_{k} ) $  such that:
  \begin{eqnarray}
    d u_{j k} \; &=& \; A_{j} - A_{k} \; \; if \, U_{j} \cap U_{k} \neq \emptyset \\
    \frac{1}{2 \pi} ( u_{j k} + u_{k l} + u_{l j} ) \; & \in & \; \mathbb{Z} \; \; if \, U_{j} \cap U_{k}  \cap U_{l} \neq \emptyset
  \end{eqnarray}
  \end{itemize}
  Posed:
  \begin{equation}
    t_{j k} \; := \; \exp ( i u_{j k} )
  \end{equation}
  it follows that:
  \begin{equation}\label{eq:first auxiliary equation-a}
    \frac{ d t_{j k}}{t_{j k}}  \; = \; A_{j} - A_{k} \; \; if \;  U_{j} \cap U_{k} \neq \emptyset
  \end{equation}
  \begin{equation}\label{eq:first auxiliary equation-b}
    t_{j k}   t_{k l}  t_{l j} \; = \; 1 \; \; if \;  U_{j} \cap U_{k}  \cap U_{l} \neq \emptyset
  \end{equation}

  By construction the maps $ \{ t_{j k} \} $ obey, furthermore, the condition:
  \begin{equation} \label{eq:second auxiliary equation}
     t_{j k}   t_{k j} \; = \; 1  \; \; if \; U_{j} \cap U_{k} \neq \emptyset
  \end{equation}
  The equation \ref{eq:second auxiliary equation} and the equation \ref{eq:first auxiliary equation-b} are equivalent to the condition that the $ \{ t_{i j} \} $ define a 1-cocycle in the $ 1^{th} $ Cech cohomology group taking values in the complex plane (thought as a group with respect to multiplication) and are a necessary and sufficient condition in order that the  $ \{ t_{i j} \} $ are the transition functions of a linear bundle $ L \stackrel{\pi}{\rightarrow}  M $.

  The equations \ref{eq:first auxiliary equation-a} and \ref{eq:first auxiliary equation-b} are furthermore a sufficient condition in order that the local potentials $ \{ A_{i} \} $ determinate a connection $ \nabla $ on such a bundle having curvature $ F_{\nabla} \;  := \; \nabla \circ \nabla  \; = \; B $.

  Since, furthermore, the $ \{ A_{i} \} $ are real 1-forms and the $ \{ t_{i j} \} $ have unitary modulus it follows that there exists an hermitian structure over $ L \stackrel{\pi}{\rightarrow}  M $ with respect to which $ \nabla $ is an hermitian connection.

  Hence $ ( ( M ,g) , B ) $ is quantizable.

  \item Let us prove that if $ ( ( M , g) , B ) $ is quantizable then the $ 2^{th} $ Weyl integrality condition holds.

  By hypothesis there exists an hermitian linear bundle $ L \stackrel{\pi}{\rightarrow} M $  endowed with an hermitian connection $ \nabla $ on it such that $ F_{\nabla} \; := \; \nabla \circ \nabla \; = \; B $.

  Let us then consider the transition functions $ \{ t_{i j} \} $ of  $ L \stackrel{\pi}{\rightarrow} M $ relative to a contractible open covering $ \{ U_{i} \} $ of M.

  Introduced the following quantities:
  \begin{equation}
     z_{j k l } \; := \; \frac{1}{2 \pi i} ( \log t_{j k} + \log t_{k l} + \log t_{l j} ) \; \; if \; U_{j} \cap U_{k} \cap U_{l} \neq \emptyset
  \end{equation}
  it follows that
  \begin{equation} \label{eq:third auxiliary equation}
    z_{j k l } \; \in \; \mathbb{Z}
  \end{equation}
  since the transition functions of a linear bundle obey the cocycle condition:
  \begin{equation}
    t_{j k } t_{k l } t_{l j} \; = \; 1
  \end{equation}
  Let us then consider the Chern class of $ L \stackrel{\pi}{\rightarrow} M  $ that, since $ F_{\nabla} \; := \; \nabla \circ \nabla \; = \; B $, is:
  \begin{equation}
    c ( F_{\nabla} ) \; = \; [ \frac{B}{2 \pi} ]
  \end{equation}

  Owing to the isomorphism existing between the cohomology groups and the Cech cohomology groups with real coefficients, to $  c ( F_{\nabla} )  $ it does corresponds a 2-cocycle of Cech cohomology with integer coefficients of which $ 2 \pi z $ is a representing element.

  Since, according to the equation \ref{eq:third auxiliary equation}, such a cocycle is at integer coefficients, it follows that:
  \begin{equation}
    c ( F_{\nabla} ) \; \in \; H^{2} ( M , \mathbb{Z} )
  \end{equation}
  \end{enumerate}
\end{proof}
 \newpage
\section{Prequantizations of a symplectic manifold}

Let us recall some basic notions concerning Souriau-Kostant's Geometric Quantization \cite{Woodhouse-91}, \cite{Souriau-97}:
\begin{definition}
\end{definition}
\emph{symplectic manifold:}

 a couple $ ( M , \omega ) $ such that:
 \begin{itemize}
   \item M is a differentiable manifold
   \item $ \omega \in Z^{2}( M ) $ is non degenerate
 \end{itemize}

 Given a symplectic manifold $ ( M , \omega) $:
 \begin{definition}
\end{definition}
\emph{prequantization of $ ( M , \omega)$}

 a couple $ (  L \stackrel{\pi}{\rightarrow} M \, , \, \nabla) $ such that:
 \begin{itemize}
   \item $ L \stackrel{\pi}{\rightarrow} M $ is an hermitian line bundle
   \item $ \nabla $ is a connection over $ L \stackrel{\pi}{\rightarrow} M $ whose curvature is $ \omega $
 \end{itemize}

 Given a symplectic manifold $ ( M , \omega ) $:
\begin{definition}
\end{definition}
\emph{ $ ( M , \omega ) $ is quantizable: }
\begin{center}
    there exist prequantizations of  $ ( M , \omega ) $
\end{center}

Let us now introduce the following:
\begin{definition} \label{def:ansatz from magnetic bottles to symplectic manifolds}
\end{definition}
\emph{Ansatz from magnetic bottles to symplectic manifolds:}
\begin{center}
    configuration space of a magnetic bottle $ \rightarrow $  phase space of a symplectic manifold
\end{center}
\begin{center}
    magnetic 2-form of a magnetic bottle $ \rightarrow $ symplectic form of a symplectic manifold
\end{center}

\bigskip

\begin{remark} \label{rem:mathematical coincidence}
\end{remark}
Let us remark that the underlying manifold of a magnetic bottle is the configuration space of dynamical systems while the underlying manifold of a symplectic manifold is the phase space of dynamical systems.

Furthermore the symplectic form of a symplectic manifold is not only closed but also non degenerate.

Last but not least a magnetic bottle is also endowed with a riemannian structure.

Anyway the non-degeneration's condition of the symplectic form plays no role in the prequantization of a symplectic manifold.

In a similar manner the riemannian structure of a magnetic bottle plays no role in the construction of an hermitian linear bundle endowed with an hermitian connection having  as curvature the magnetic 2-form.

It follows that, from a mathematical point of view, the prequantization of a symplectic manifold and the construction of the quantum Hilbert space of states of a magnetic bottle are mathematically identical procedures.

\bigskip

\begin{theorem} \label{th:theorem of the second Weyl integrality condition for a symplectic manifold}
\end{theorem}
\emph{Theorem about $ 2^{th} $ Weyl integrality condition:}
\begin{center}
    $ ( M , \omega ) $ is quantizable $ \Leftrightarrow \; [ \frac{\omega}{2 \pi} ] \in H^{2} (M , \mathbb{Z} ) $
\end{center}
\newpage
\begin{proof}
It is sufficient to restate the proof of the theorem \ref{th:theorem on the second Weyl integrality condition} making the ansatz
\ref{def:ansatz from magnetic bottles to symplectic manifolds} and taking into account the remark \ref{rem:mathematical coincidence}.
\end{proof}

\bigskip

\begin{theorem} \label{th:multiplicity of prequantization}
\end{theorem}
\emph{Multiplicity of prequantizations:}
\begin{center}
    the set of inequivalent prequantizations of $  ( M , \omega ) $ is parametrized by $ Hom ( H_{1} ( M , \mathbb{Z}) , U(1)) $
\end{center}
\begin{proof}
Applying the inverse of the ansatz \ref{def:ansatz from magnetic bottles to symplectic manifolds} (taking into account the remark \ref{rem:mathematical coincidence}) to the first part of the proof of the theorem \ref{th:theorem on the second Weyl integrality condition} it follows that an hermitian linear bundle $  \; L \stackrel{\pi}{\rightarrow} M $  endowed with an hermitian connection $ \nabla $ on it such that $ F_{\nabla} \; := \; \nabla \circ \nabla \; = \; \omega $ is determined by a triple $ ( \{ U _{i} \} , \{ A_{i} \} , \{ u_{j k} \})  $ such that:
  \begin{itemize}
    \item $ \{ U_{i} \} $ is a contractible open cover of M
    \item $ \{ A_{i} \} $ is a collection of local potentials of the magnetic field:
    \begin{equation}
        A_{i} \in \Omega^{1}( U_{i} ) \; : \; B|_{U_{i}} \, =  \, d A_{i}
    \end{equation}

  \item $ \{ u_{j k} \} $ is a collection of maps $ u_{j k} \in C^{\infty} ( U_{j} \cap U_{k} ) $  such that:
  \begin{equation}  \label{eq:fourth auxiliary equation-a}
     d u_{j k} \; = \; A_{j} - A_{k} \; \; if \, U_{j} \cap U_{k} \neq \emptyset
  \end{equation}
  \begin{equation} \label{eq:fourth auxiliary equation-b}
    \frac{1}{2 \pi} ( u_{j k} + u_{k l} + u_{l j} ) \; \in \; \mathbb{Z} \; \; if \, U_{j} \cap U_{k}  \cap U_{l} \neq \emptyset
  \end{equation}

  \end{itemize}

By performing on such a triple  $ ( \{ U _{i} \} , \{ A_{i} \} , \{ u_{j k} \})  $  the following ansatz:
\begin{equation}
    u_{j k} \; \mapsto  \; u'_{j k}  \, :=  \, u_{j k} +  y_{j k}
\end{equation}
where $ y_{j k} \in \mathbb{R} $ is such that:
\begin{eqnarray}
  y_{j k} \; &=& \; - y_{k j} \\
  \frac{1}{2 \pi} (  y_{j k} +  y_{k l} +  y_{l j})  \; & \in & \; \mathbb{Z} \; \; if \; U_{j}\cap U_{k} \cap U_{l} \neq \emptyset
\end{eqnarray}
the new triple $ ( \{ U _{i} \} , \{ A_{i} \} , \{ u'_{j k} \})  $ satisfies again the conditions of the equation \ref{eq:fourth auxiliary equation-a} and the equation \ref{eq:fourth auxiliary equation-b}.

Such an ansatz corresponds, therefore, to replace the fibre bundle $  \; L \stackrel{\pi}{\rightarrow} M $ with the fibre bundle  $  \; L \otimes F \stackrel{\pi}{\rightarrow} M $ where
$   \; F \stackrel{\pi}{\rightarrow} M $ is the hermitian linear bundle having the following transitions functions:
\begin{equation}
    t_{j k} \: := \; \exp ( i y_{j k} )
\end{equation}
Since the $ t_{j k} \in U(1) $ are constants it follows that the fibre bundle $   F \stackrel{\pi}{\rightarrow} M $ is flatizable, i.e. it admits a connection with null curvature.

Hence $  \; L \otimes F \stackrel{\pi}{\rightarrow} M $ admits a connection $ \nabla' $ such that $ F_{\nabla'} \; := \; \nabla' \circ \nabla' \;  =  \; B $.

As a result, given an hermitian linear bundle $  \; L \stackrel{\pi}{\rightarrow} M $  endowed with the hermitian connection $ \nabla $ on it such that $ F_{\nabla} \; := \; \nabla \circ \nabla \; = \; \omega $, other hermitian linear bundles endowed with the hermitian connection $ \nabla $  having as curvature $ \omega $ may be obtained multipliying tensorially for flatizable hermitian linear bundles.

That all the hermitian linear bundles endowed with an hermitian connection $ \nabla $ having as curvature $ \omega $ may be obtained in this way, furthermore, follows by observing that given an hermitian linear bundle $  \; L_{1} \stackrel{\pi}{\rightarrow} M $  endowed with the hermitian connection $ \nabla_{1} $ on it such that $ F_{\nabla_{1}} \; := \; \nabla_{1} \circ \nabla_{1} \; = \; \omega $ and another  hermitian linear bundle $  \; L_{2} \stackrel{\pi}{\rightarrow} M $  endowed with the hermitian connection $ \nabla_{2} $ on it such that $ F_{\nabla_{2}} \; := \; \nabla_{2} \circ \nabla_{2} \; = \; \omega $ the fibre bundle $   L_{1}^{- 1} \otimes L_{2} \stackrel{\pi}{\rightarrow} M $ is flatizable.

Hence the set of possible prequantizations of $ (M , \omega )$ are in bijective correspondence with the set of the the flatizable hermitian linear bundles over M.

From the other side, as it can be appreciated considering the restriction posed on the transition functions seen as cocycles of Cech cohomology, these are in bijective correspondence with the elements of $ H^{1}( M , U(1)) $.

Since:
\begin{equation}
    H^{1}( M , U(1)) \; = \; Hom ( H_{1} ( M , \mathbb{Z}) , U(1))
\end{equation}
the thesis follows.
\end{proof}

\bigskip

\begin{remark}
\end{remark}
In the framework of Geometric Quantization the prequantization of the phase space is not the whole story since, after it, the procedure of quantization is performed by restricting the Hilbert space of states, so that the wave function doesn't depend on both position and momentum, by introducing a polarization, i.e a sort of fibration of the phase space in lagrangian submanifolds of maximal dimension.
\newpage
\section{Quantization of a non-trivial multiply-connected magnetic bottle: multiplicity of quantizations}
Given a non-trivial multiply-connected magnetic bottle $ ( ( M , g) , B ) $ the mathematical analogy between Yves Colin de Verdiere's quantization formalism for magnetic bottles and the formalism of prequantization in the framework of Souriau-Kostant's Geometric Quantization may be used to infer that:
\begin{theorem} \label{th:multiplicity of quantization}
\end{theorem}
\emph{Multiplicity of quantizations:}
\begin{center}
    the set of inequivalent quantizations of a magnetic bottle  $  (( M , g)  , B) $ is parametrized by $ Hom ( H_{1} ( M , \mathbb{Z}) , U(1)) $
\end{center}
\begin{proof}
It is sufficient to restate the proof of the theorem \ref{th:multiplicity of prequantization} making the inverse of the ansatz \ref{def:ansatz from magnetic bottles to symplectic manifolds} (taking into account the remark \ref{rem:mathematical coincidence}) to obtain the thesis.
\end{proof}

\bigskip

The multiplicity of quantizations stated by the theorem \ref{th:multiplicity of quantization} is nothing but a manifestation of the Homotopy Superselection Rule, different quantizations corresponding to different homotopy superselection sectors.

Hence the quantum hamiltonian of the quantum dynamical system obtained quantizing the magnetic bottle $  (( M , g)  , B) $ has quantum hamiltonian:
\begin{equation}
    ( D( \hat{H}) , \hat{H} ) \; = \; \otimes_{\theta \in  Hom ( H_{1} ( M , \mathbb{Z} ) , U(1)) } ( D( \hat{H}_{\theta}) , \hat{H}_{\theta} )
\end{equation}
where:
\begin{equation}
  ( D( \hat{H}_{\theta}) , \hat{H}_{\theta} ) \; := \; K [ \overline{ ( D(E) := \Gamma_{0}^{\infty}( L_{\theta}) \, , \, E ( \psi_{1} , \psi_{2}) := \int_{M} d \mu_{g} < \nabla_{\theta} \psi_{1} , \nabla_{\theta} \psi_{2} >_{\theta}  )} ]
\end{equation}
where  $  L_{\theta} \stackrel{\pi}{\rightarrow} M $  is the hermitian linear bundle over M endowed with the hermitian connection $ \nabla_{\theta} $ on it such that $ F_{\nabla_{\theta}} \; := \; \nabla_{\theta} \circ \nabla_{\theta} \; = \; B $ parametrized by the homotopic superselection charge $ \theta \in  Hom ( H_{1} ( M , \mathbb{Z} ) , U(1)) $.

\bigskip

\begin{example}
\end{example}
Let us consider in particular the case in which $ M = \Sigma_{g} $ is a Riemann surface of genus $ g \neq 0 $.

Then the quantum hamiltonian may be expressed as:
\begin{equation}
    ( D( \hat{H}) , \hat{H} ) \; = \; \int_{\Phi}^{\otimes} d \vec{\theta} \, ( D( \hat{H}_{\vec{\theta}}) , \hat{H}_{\vec{\theta}} )
\end{equation}
where :
\begin{equation}
  ( D( \hat{H}_{\vec{\theta}}) , \hat{H}_{\vec{\theta}} ) \; := \; K [ \overline{ ( D(E) := \Gamma_{0}^{\infty}( L_{\theta}) \, , \, E ( \psi_{1} , \psi_{2}) := \int_{\Sigma_{g}} d \mu_{g} < \nabla_{\vec{\theta}} \psi_{1} , \nabla_{\vec{\theta}} \psi_{2} >_{\vec{\theta}}))   } ]
\end{equation}
where  $  L_{\vec{\theta}} \stackrel{\pi}{\rightarrow} \Sigma_{g} $  is the hermitian linear bundle over $ \Sigma_{g}$ endowed with the hermitian connection $ \nabla_{\vec{\theta}} $ on it such that $ F_{\nabla_{\vec{\theta}}} \; := \; \nabla_{\vec{\theta}} \circ \nabla_{\vec{\theta}} \; = \; B $ parametrized by the homotopic superselection charge $ \vec{\theta} $ belonging to the flux torus $ \Phi = T^{2 g} $.
\newpage
\section{Notation}
\begin{center}
\begin{tabular}{|c|c|}
  \hline
  $ [ G , G ] $ & commutator subgroup of the group G \\
  $ S^{k}(M) $ & $k^{th}$ Sobolev space of the manifold M \\
  $ < \cdot , \cdot >_{g} $ & inner product induced by the riemannian metric g \\
  $ d \mu_{g} $ & invariant measure on the riemannian manifold (M , g) \\
  $ \nabla_{g} $ & Levi-Civita connection of the riemannian manifold (M , g)  \\
  $\Delta_{g}$ & Laplace-Beltrami operator of the riemannian manifold (M , g)  \\
  $ R_{g} $ & scalar curvature of the riemannian metric g \\
  $ Ntf( \Sigma_{g}) $ & normal topological form of the Riemann surface $ \Sigma_{g} $ \\
  $ Hom(M_{1}, M_{2})$ & homeomorphisms between the topological spaces $ M_{1}$ and $ M_{2}$ \\
  $ \Omega^{n}(M) $ & n-forms over M \\
  $Z_{n}(M,G)$ & n-cycles over M with respect to G \\
  $B_{n}(M,G)$ & n-boundaries over M with respect to G \\
  $H_{n}(M,G)$ & $ n^{th}$ homology group over M with respect to G \\
  $Z^{n}(M,G)$ &  n-cocycles over M with respect to G \\
  $B^{n}(M,G)$ & n-coboundaries over M with respect to G \\
  $H^{n}(M,G)$ & $ n^{th}$ cohomology group over M with respect to G \\
  $ LOOPS(M) $ & loops over M \\
  $ \pi_{n}(M)$ & $ n^{th}$ homotopy group of M \\
  $ c(F) $  & Chern class of a fibre bundle endowed with a connection having curvature F \\
  $ C^{\infty}(M)$ & smooth functions over M \\
  $ C_{0}^{\infty}(M)$ & smooth functions with compact support over M \\
  $ \Gamma^{\infty}(L)$ & smooth sections of the fibre bundle L \\
  $ \Gamma_{0}^{\infty}(L)$ & smooth sections  with compact support of the fibre bundle L \\
  $ \Gamma^{\infty}(L,U)$ & smooth local sections over the open U of the fibre bundle L \\
  $ \Gamma_{0}^{\infty}(L,U)$ & smooth local sections with compact support over the open U of the fibre bundle L \\
  $ Hol_{\gamma}( \nabla ) $ & holonomy of the connection $ \nabla $ along the loop $ \gamma $ \\
  $ \overline{(D(E),E)} $ & closure of the symmetric form $ (D(E),E) $ \\
  $ K [ (D(E),E) ] $ & Kato operator of the closed symmetric form $ (D(E),E) $ \\
  \hline
\end{tabular}
\end{center}

\end{document}